\documentclass[journal]{STTemplate}
\usepackage{graphicx}
\usepackage{float}	

\begin{document}
\title{Design, Development and Comparison of two Different Measurement Devices for Time-Resolved Determination of Phase Shifts of Bioimpedances}

\author{R.~Kusche,
        S.~Kaufmann,
        and M.~Ryschka 
\thanks{R. Kusche is with the Laboratory for Medical Electronics, L\"ubeck University of Applied Sciences (telephone:+ 49 451 300 5400, e-mail:\newline romankusche@gmx.de).}
\thanks{S. Kaufmann is with the Laboratory for Medical Electronics, L\"ubeck University of Applied Sciences (telephone:+ 49 451 300 5400, e-mail: kaufmann@fh-luebeck.de).}
\thanks{M. Ryschka is with the Laboratory for Medical Electronics, L\"ubeck University of Applied Sciences (telephone:+ 49 451 300 5026, e-mail: ryschka@fh-luebeck.de).}}

\maketitle
\pagestyle{empty}
\thispagestyle{empty}

\begin{abstract}
Bioimpedance measurements are a non-invasive method to determine the composition of organic tissue. For measuring the complex bioimpedance between two electrodes, an alternating current with a constant amplitude is injected into the tissue. The developed voltage drop is used to calculate the real and imaginary part of the impedance under test. Measurements in the past indicated that it could be possible that the beating of the heart has an effect on the measured phase shift of the impedance under test. In this work a hardware system is developed, capable of measuring changes of the bioimpedance phase shift with high resolution. Two different measurement methods are used. The first method is an analog circuit, which has been developed in a previous project. The other method uses the integrated circuit AFE4300 (Texas Instruments) \cite{AFE4300}. The results of both methods are transmitted with a microcontroller (Xmega256A3BU from Atmel Corp.) via the USB interface to a host PC. On the host PC the visualization of the measurements and the control of the embedded system is achieved with a developed LabView program. The work includes the development of the hardware and software, as well as a comparison of the accuracy of the two measurement methods. The result of a verification of the systems is that both measurement methods have an effective resolution of about 0.3$^{\circ}$. 
\end{abstract}

\section{Introduction}
Bioimpedance measurements are a non-invasive method for determining the electrical characteristics of organic tissues \cite{BioimpedanceMonitoringforphysicians}. With time-resolved measurements, it is possible to determine dynamic characteristics like the pulse wave or the effect of breathing \cite{WebsterMedicalInstrumentation}.

\begin{figure}[!h]
	\centering
	\includegraphics[width = 0.25\textwidth]{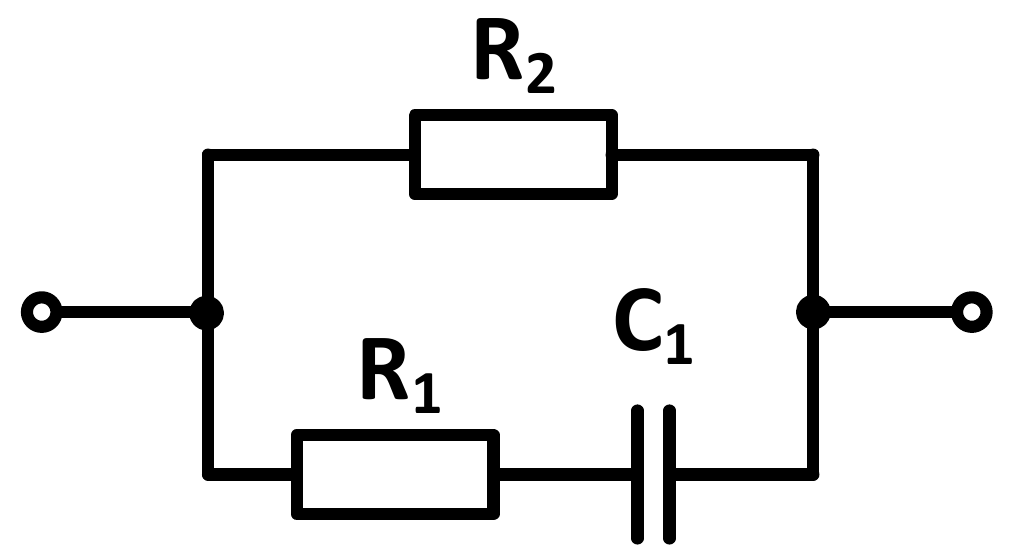}
	\caption{Simplified electrical equivalent circuit of biological tissues}
	\label{fig:esb}
\end{figure}

For measuring the magnitude and phase of the unknown bioimpedance, a small known alternating current is injected into the tissue. This current produces a voltage drop across the bioimpedance, which is measured and used to calculate the complex bioimpedance. Usually the measured impedance phase of organic tissues is negative. This behavior can be explained by the capacitive behavior caused by the cell structure of biological matter \cite{GrimnesBioimpedance}. Fig. \ref{fig:esb} shows a simplified electrical equivalent circuit, whereby $R_2$ represents the extracellular fluid and $R_1$ and $C_1$ represent the cell impedance with cell membrane and intracellular fluid.
When the bioimpedance is measured time resolved a small variation of the impedance magnitude can be observed which is synchronous with the heart beat. 

The reason for this effect is the pulsation of the blood in the tissue under test. When the heart is pumping blood through the bloodvessels, the pressure in the arteries increases abruptly and then falls again. Because of this variation of pressure, the arteries diameter and thus the volume of blood in the tissue under test changes.
Since the conductivity of blood is significantly higher than that of tissue, this leads to a change of the impedance.
One may assume that besides the magnitude also the phase of the complex bioimpedance is influenced by the blood pulsation. The changing of the bioimpedance phase shift would indicate that the blood additionally has a different phase shift than the tissue.
Typical bioimpedance phase shifts of living tissues are between $\approx-10^{\circ}$ to $0^{\circ}$, with a maximum at about $50~kHz$ \cite{BMS2}.

This work compares two different phase shift measurement methods. The first method is an analog circuit, realized with Operational Amplifiers (OPA). The second method is implemented by the IC (Integrated Circuit) AFE4300 from Texas Instruments. For the evaluation of the different measurement methods, both methods are realized together with an Electrocardiography (ECG) circuit, as a physiological time reference, on the same Printed Circuit Board (PCB). The PCB also contains a microcontroller system for Analog to Digital Conversion (ADC), basic signal preprocessing and USB for data transmission to a host PC for further signal processing and displaying of the measurement data.

\section{Measuring Methods}
\subsection{Analog Measurement System}
The basic idea of the analog phase shift measurement circuit is the subtraction of two alternating voltages with the same frequency and amplitude.
It can be shown that for small phase shifts the difference of two equal sinusoidal voltages is almost linear to their phase shift \cite{KuscheProjekt}.
To apply this method to determine the phase shift between the injection current, represented by the voltage drop across a shunt resistor and the voltage drop across the bioimpedance, both voltages have to be equalized in amplitude.
In the actual implementation, this is achieved via normalization. Afterwards the difference signal is rectified and low pass filtered to get a DC signal. The circuit has been realized within a previous project \cite{KuscheProjekt}. 

\subsection{Integrated Measuring System (AFE4300)}
The AFE4300 is an IC including analog measurement circuits as well as a SPI bus for the communication with a microcontroller. The interface between the analog and the digital part is internally realized with an ADC ($16~bit$, $868~SPS$) and a Digital to Analog Converter (DAC, $6~bit$, $1~MSPS$). Thus the IC is an integrated measurement system, there are different kinds of signal chains integrated. In this work only the body composition meter in I/Q (In-phase/Quadrature)-Demodulator mode is used. 
The necessary sinusoidal excitation current is also generated inside the integrated circuit with a Voltage Controlled Current Source (VCCS).

\section{Hardware Development}
\begin{figure}[!h]
	\centering
	\includegraphics[width = 0.5\textwidth]{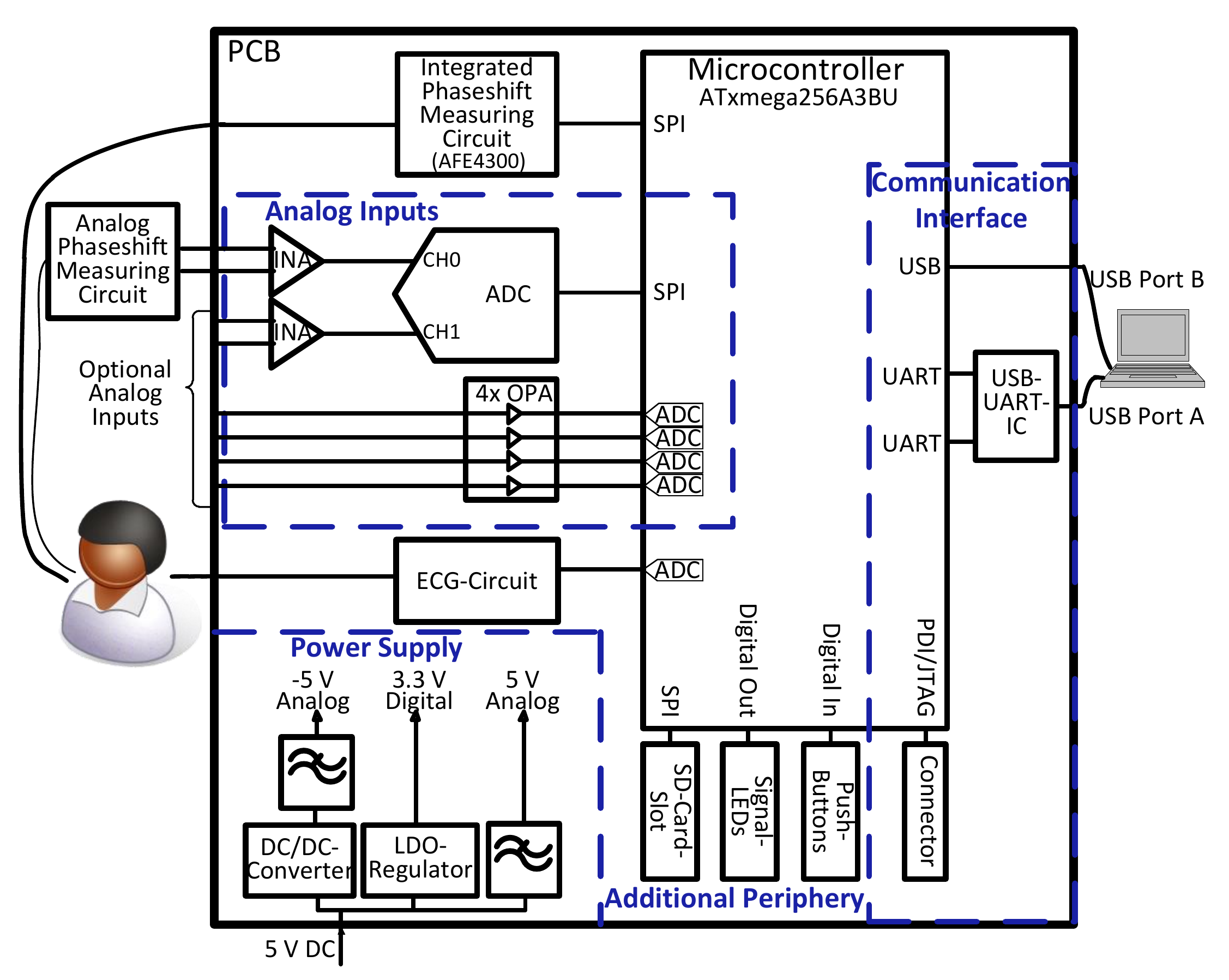}
	\caption{Block diagram of the realized PCB}
	\label{fig:bsb}
\end{figure}

The block diagram in Fig. \ref{fig:bsb} shows the schema of the developed PCB for interfacing the object under test to the host PC. For the communication between the AFE4300 and the microcontroller, a SPI bus is used. However the external analog phase shift measurement circuit uses the PCB's analog inputs.

\subsection{Analog Inputs}
To interface the analog phase shift measurement circuit an Instrumentation Amplifier (INA, LT1789-1 from Linear Technology) is used. The results are fed to one channel of a $16~bit$, $150~kSPS$ dual channel ADC (LTC1865L from Linear Technology). The second ADC channel is prepared for future use, together with four further analog inputs which digitalization is provided by the micro controller's internal $12~bit$, $2~MSPS$ ADCs. Signal buffering is achieved by voltage followers realized with OPAs (OPA4134
from Texas Instruments). The external high accuracy ADC is connected via SPI to the microcontroller.

\subsection{ECG Circuit}
For common mode reduction, the ECG module is equipped with a
Driven Right Leg (DRL) circuit and active driven shields. It is based on an INA circuit (LT1789-1 from Linear Technology) with baseline wandering rejection. In order to reduce noise an optional active 2\textsuperscript{nd} order lowpass filter and a passive $50~Hz$ Twin-T notch filter are implemented \cite{Halbleiterschaltungstechnik}.

\subsection{Power Supply}
The system is externally supplied with $5~V_{DC}$, which could be provided from the USB. Though for electrical safety considerations it is recommended to power the embedded system with an IEC-60601-1 compliant power source and to isolate the USB via a fiber optic USB hub. Internally the system needs $\pm5.0~V$ for the analog parts and $3.3~V$ for the digital components.  
Whereas the $3.3~V$ are generated via a LDO-regulator (LT1129 from Linear Technology), the $-5.0~V$ are generated via a DC/DC-Converter (LT3479 from Linear Technology). The converter's switching frequency is kept above $1.5~MHz$ to be far away from the measurement frequency ranges. The positive as well as the negative supply voltages are filtered by analog low pass filters with a cut off frequency of $\approx7~kHz$.

\subsection{Microcontroller System and Communication Interfaces}
The used microcontroller (ATxMega256A3BU from Atmel Corp.) is an $8~bit$ controller with $256~kB$ of program memory. It was chosen based on its peripheral features like integrated ADCs and the USB-interface. The microcontroller has a clock frequency of $32~MHz$. For future use, further hardware components are implemented on the PCB like a SD-Card slot, signal LEDs and push buttons.

Fig. \ref{fig:bsb} shows that the system has two USB ports (Port A \& Port B). The difference between them is that Port B directly connects the microcontroller with the PC. This is the standard port for measurements. Port A connects the microcontroller and the PC via an interface IC (FT2232D from Future Technology Devices Inc.). Port A is only used for debug purposes. It is possible to program the microcontroller via the PDI or the JTAG interface also.

\begin{figure}[!h]
	\centering
	\includegraphics[width = 0.48\textwidth]{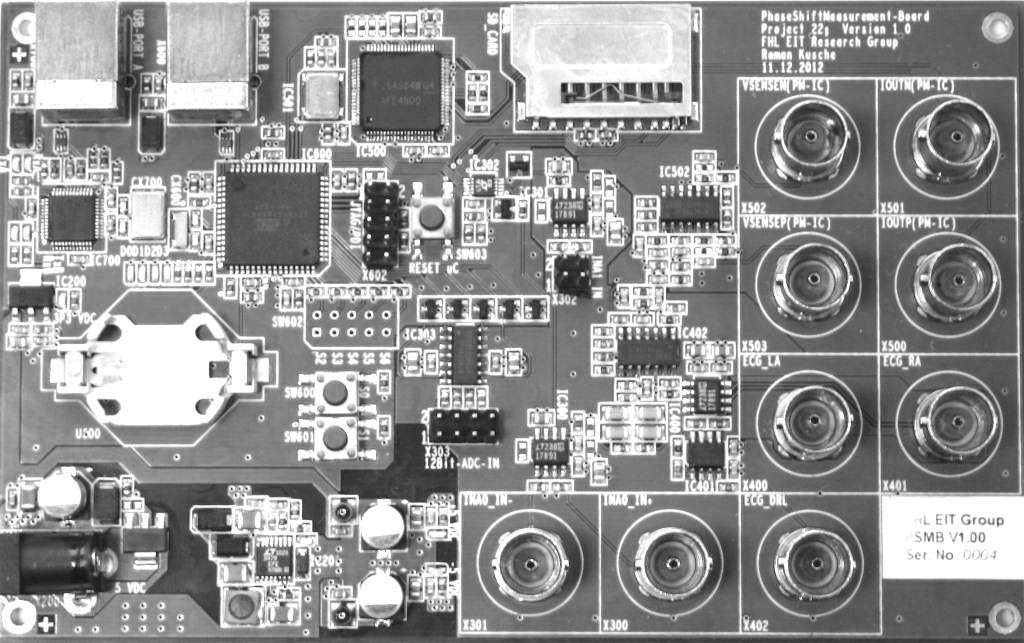}
	\caption{Manufactured phase shift measurement PCB.}
	\label{platine}
\end{figure}

Fig. \ref{platine} shows the manufactured and populated four layer PCB of the measurement system. It contains more than 300 components and has dimensions of about $150~mm$ x $94~mm$.

\section{Software Development}
The development of the software can be separated into two parts. The first part is programming the microcontroller's firmware. In the second part the LabView Software for the PC is programmed.

\subsection{Microcontroller Firmware}
\begin{figure}[!h]
	\centering
	\includegraphics[width = 0.48\textwidth]{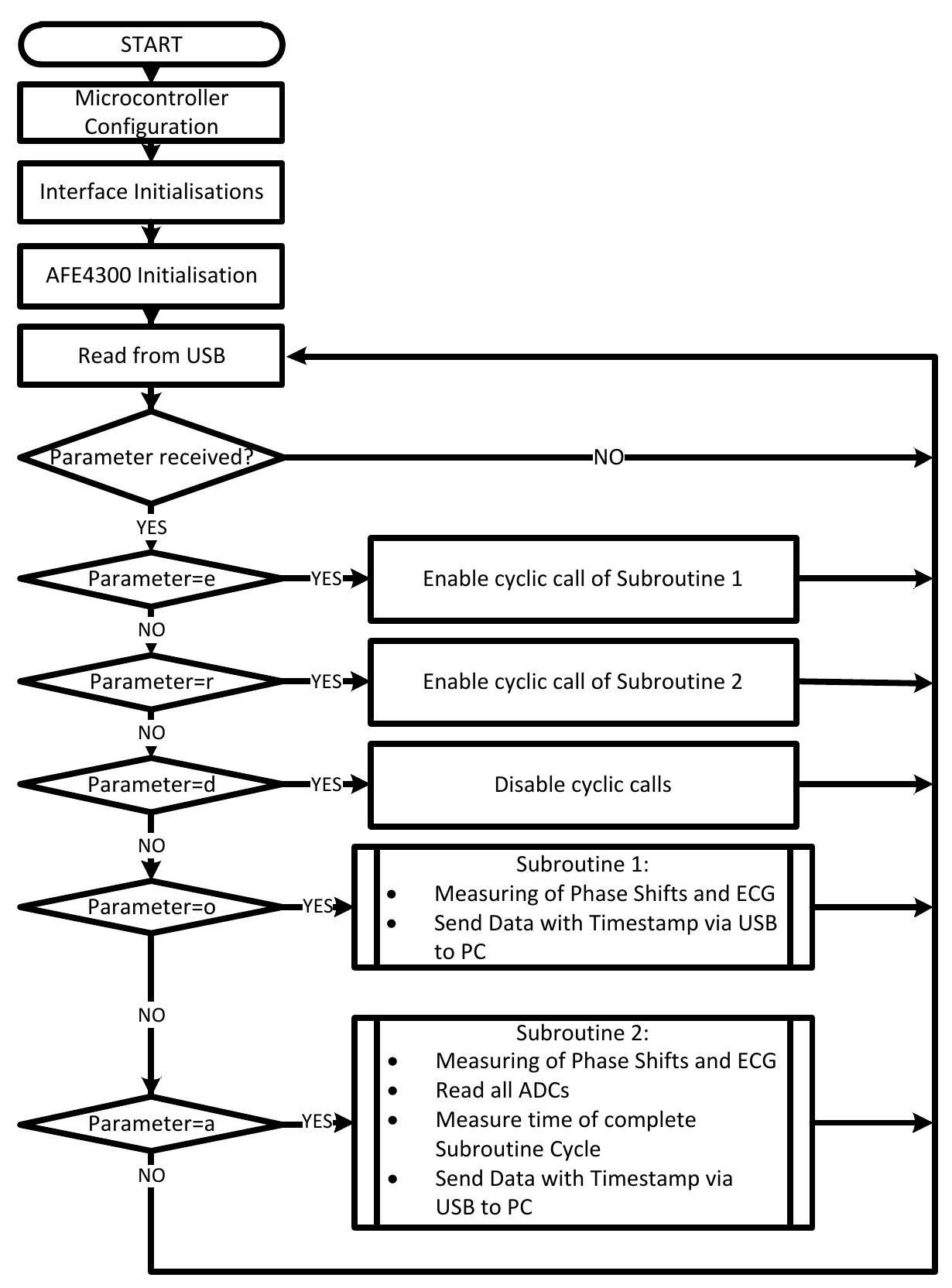}
	\caption{Simplified flow chart of the developed microcontroller firmware.}
	\label{controller}
\end{figure}

For the firmware development the Atmel Software Framework (ASF) is used as an aid. Configuring the microcontroller's USB-Port as a Virtual Serial Port simplified the communication with the PC. Fig. \ref{controller} shows the flow chart of the firmware.
The first steps execute the hardware configurations and initializations. In the main loop the USB-Interface is checked for new received data. The data packet sent from the PC consists of only one character. After a case distinction the respective measuring mode is executed. Subroutine 1 just measures phase shifts and sends the results to the PC. Subroutine 2 additionally measures all the other analog inputs. So Subroutine 1 is faster and the measuring cycle is shorter ($T_{Sub1} \approx 3.2~ms$, $T_{Sub2} \approx 4.5~ms$).

\subsection{Labview Software}
\begin{figure}[!h]
	\centering
	\includegraphics[width = 0.48\textwidth]{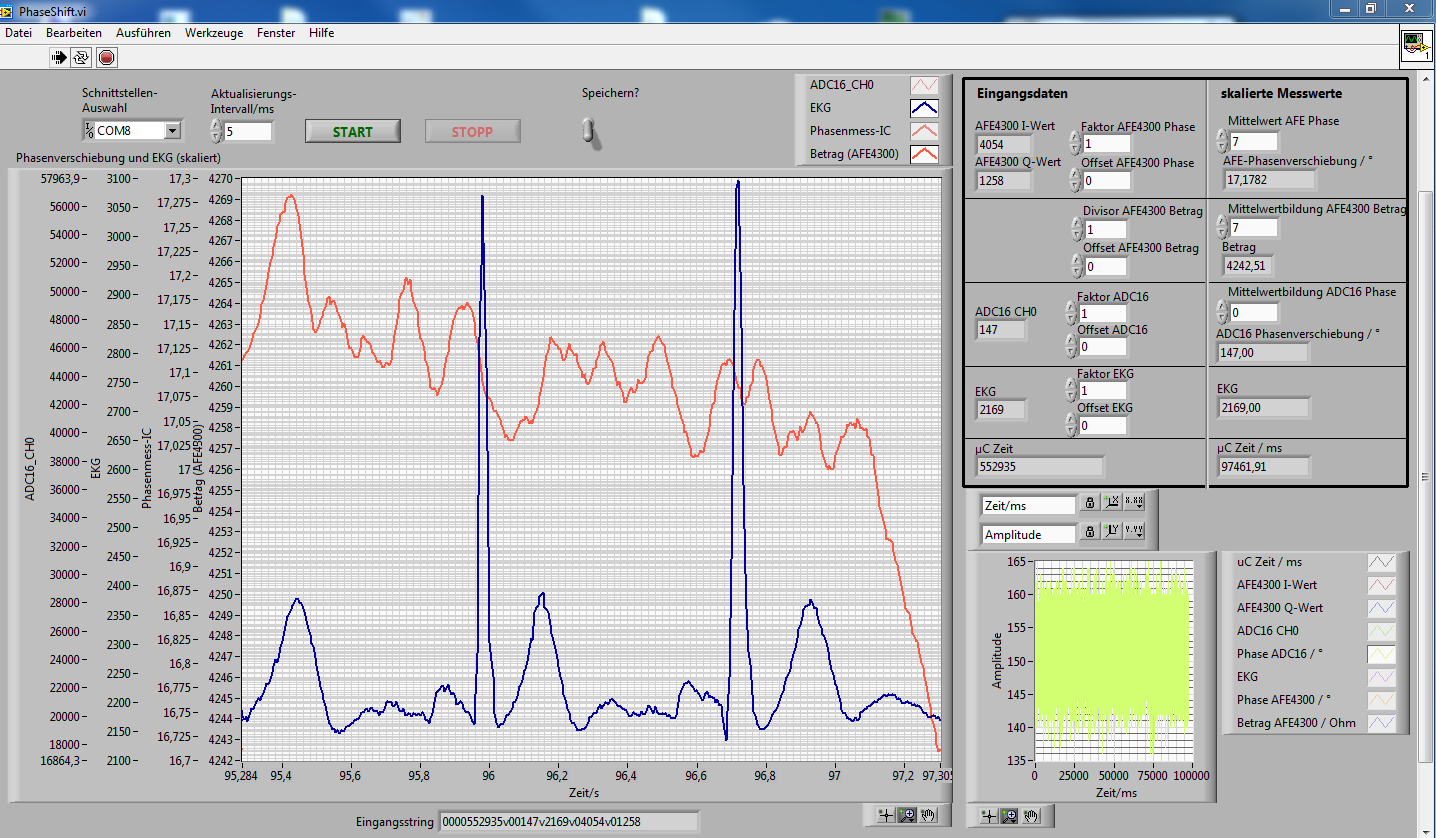}
	\caption{Screenshot of the developed LabView software.}
	\label{screenshot}
\end{figure}

The programmed LabView software's purpose is displaying the measurements in real-time. It is also possible to change the sample rates and activate or deactivate the ADC-channels. Additionally the results can be exported to an MATLAB or Excel compatible data format. To reduce the influence of noise a moving average filter can be activated. In Fig. \ref{screenshot} a screenshot of the software's user interface is shown.

\section{Verification and Measurements}
\begin{figure}[!h]
	\centering
	\includegraphics[width = 0.25\textwidth]{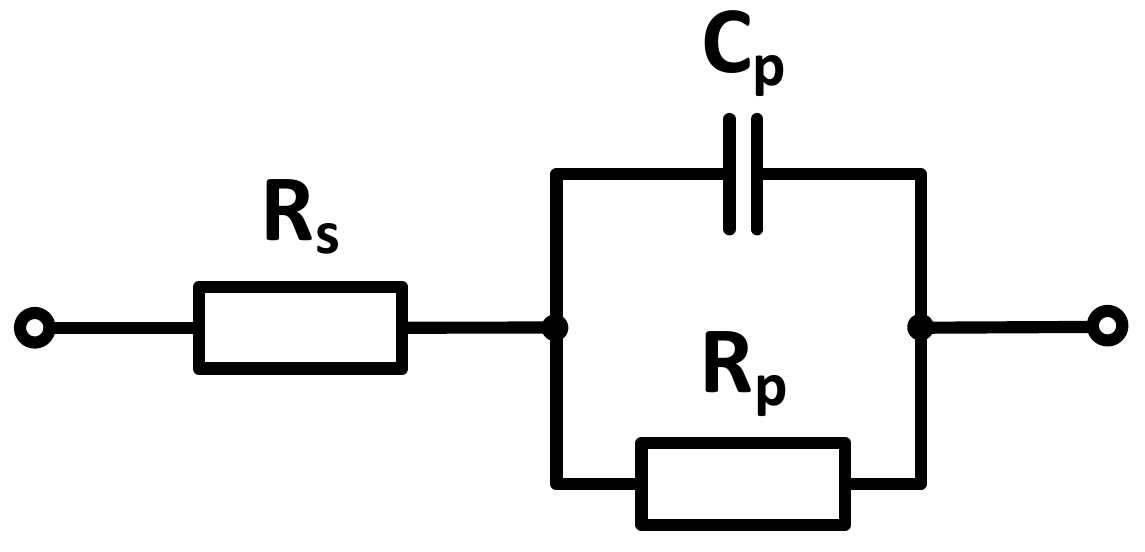}
	\caption{Equivalent circuit for bioimpedances used for verification}
	\label{fig:rcr}
\end{figure}

For the verification of the phase shift measurements an equivalent circuit for bioimpedances according to Fig. \ref{fig:rcr} is used. The result of an error estimation, executed with the total differential, is that this circuit is more insensitive to the components tolerances than the circuit shown in Fig. \ref{fig:esb}.

To realize the desired magnitudes and phase shifts only the resistors were changed. 
The reason is the higher tolerance of the capacitance. It is 1\% whereas the resistors tolerance is just 0.1\%. An additional advantage is that resistors are available in a wide range of values.
The verification of both systems was executed with a frequency of $62.5~kHz$.
An error estimation resulted to a maximal phase shift error of the equivalent circuit of less than $\pm0.1^{\circ}$ in the interesting range.

\begin{figure}[!h]
	\centering
	\includegraphics[width = 0.48\textwidth]{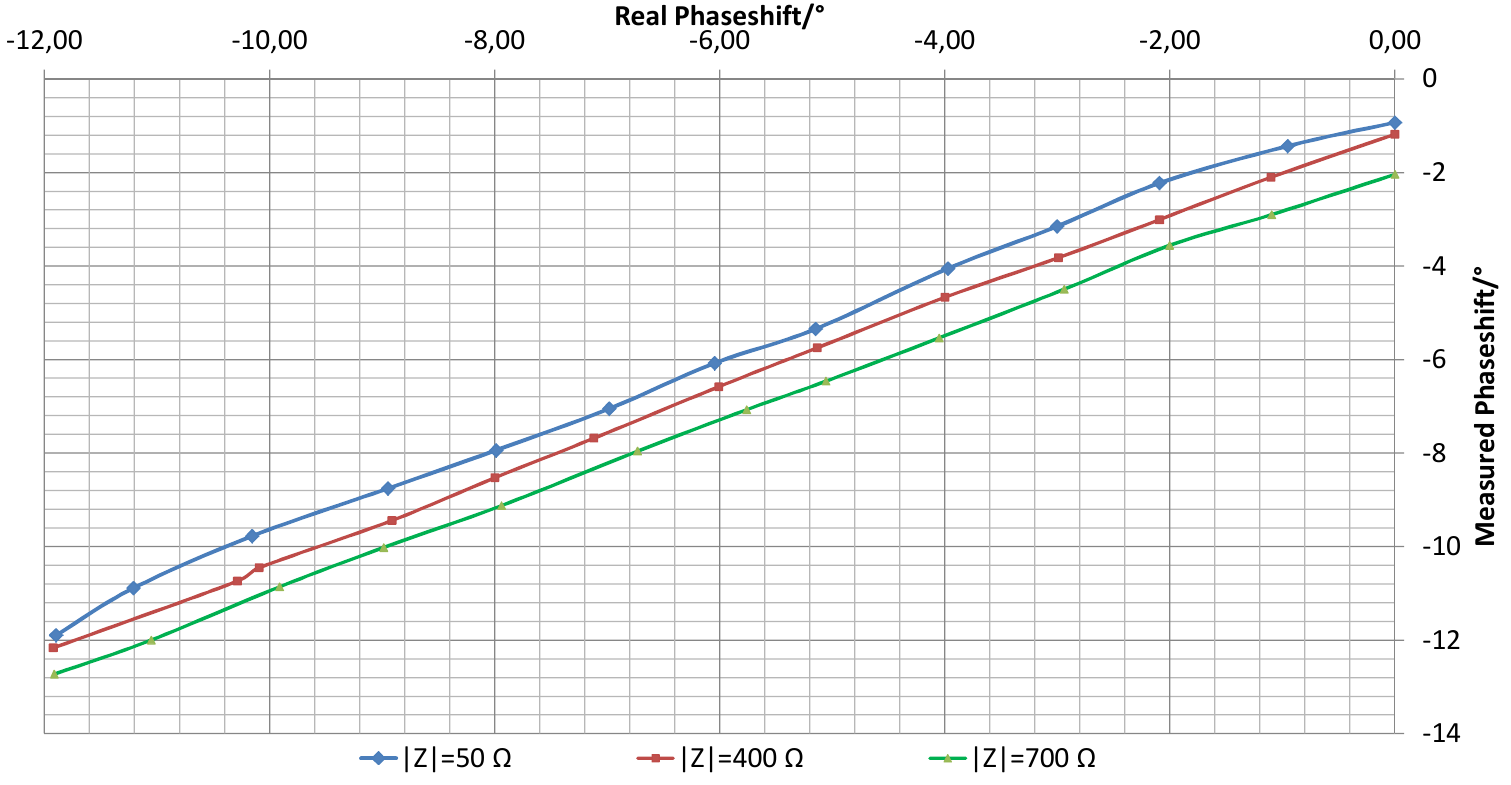}
	\caption{Measuring results of the analog system. The phantom's phase shift was varied in a range of $0^{\circ}$ to $-12^{\circ}$. Verification executed for magnitudes of $50~\Omega$, $400~\Omega$ and $700~\Omega$.}
	\label{analog}
\end{figure}

\begin{figure}[!h]
	\centering
	\includegraphics[width = 0.48\textwidth]{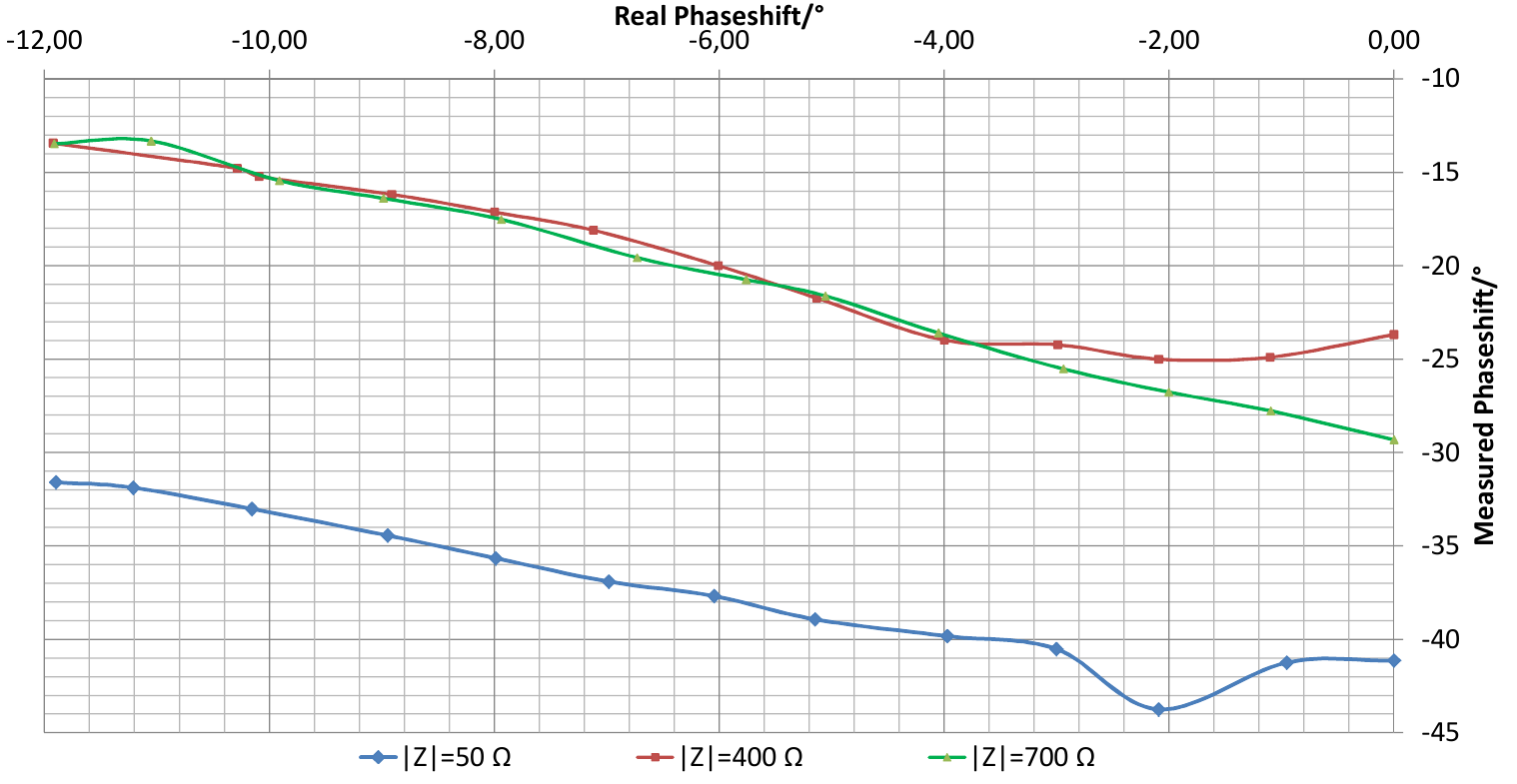}
	\caption{Measurement results of the AFE4300. The phantom's phase shift was varied in a range of $0^{\circ}$ to $-12^{\circ}$. Verification executed for magnitudes of $50~\Omega$, $400~\Omega$ and $700~\Omega$.}
	\label{ic}
\end{figure}

The impedances were realized on a separate PCB where switches are used to change the resistances. Fig. \ref{analog} shows the result of the analog phase shift measurement system. Achieving this characteristic was only possible by adapting the system's shunt resistance to the magnitude of the measured impedance. This measurement system shows a good linearity but some offset depending on the magnitude of the measured impedance.
Fig. \ref{ic} shows the results of the AFE4300. The measurements depend on the impedance's magnitude, too. Furthermore the results have large offsets and are inverted.
Especially for small phase shifts below $4^{\circ}$ there are strong deviations from a linear relation.
\section{Results and Discussion}
The linear behavior of the analog measurement system, especially when there are small phase shifts, is an advantage. However the system's major disadvantage is that an external current source is needed.
For suitable results it is also important to adjust the shunt resistor. As a consequence the magnitude of the impedance under test has to be known.
The IC does not have such a linear behavior as the analog circuit. Although up to four calibration impedances can be connected to the IC pins.
With these known impedances and a microcontroller program it is possible to automate the calibration.
The disadvantage of using an IC with such a small housing (LQFP80) is the effort of implementation for experimental purposes. A complete PCB with a microcontroller has to be developed and manufactured. 
However there is an essential advantage that the whole measurement is done by just one component. Thus no external current source and shunt resistor is needed. Furthermore the AFE4300 is low priced (2.50 \$ per 1000 pcs.) and no additional ADC is required.
For reliable measuring of absolute phase shifts both systems have to be calibrated comprehensively.
The effective resolution of both systems is about $0.3^{\circ}$.
Measurements of real bioimpedances could not exhibit a relationship between the pulse and the phase shift.
Using another impedance measurement instrument, which has also been developed in the work group \cite{AhighaccuracyBMS, MESI},
it could be shown that the blood pulsation induced phase shift changes are only in a range of about $0.01^{\circ}$.
Hence for this application both the analog and the integrated phase shift measurement systems are inappropriate.

\section{Conclusions}
For time depending measuring of the bioimpedances phase shift, two different measurement systems were compared.

The contrasted systems are the integrated circuit AFE4300 from Texas Instruments and an analog phase shift measurement circuit realized on a separate board.

To compare both systems a microcontroller based hardware system was developed and manufactured.
Additionally the resulting PCB is able to measure an ECG simultaneously. 
A firmware was programmed for sending the measurement data from the microcontroller via the USB interface to a PC.
Furthermore a LabView software was developed for signal processing and data logging.
The systems were verified with the help of a RC-phantom with known magnitudes and phases.
The result of the verification was that both circuits do have advantages and disadvantages in implementation as well as in the measuring results.
But the effective phase shift resolution of both systems is far to low to detect the pulse induced phase variation of living tissue.
With the developed PCB it is possible to test further measurement methods in the future.


\begin{thebibliography}{1}
	\bibitem{AFE4300}
	Texas Instruments. AFE4300, Datasheet.

	\bibitem{BioimpedanceMonitoringforphysicians}
	A. Ivorra, Bioimpedance Monitoring for physicians: an overview.  Centre Nacional de	Microelectrònica - Biomedical Applications Group, 2003.
	
	\bibitem{WebsterMedicalInstrumentation}
	J. G. Webster, Medical Instrumentation- Application and Design, 4th ed. Wiley, 2010. 
	
	\bibitem{GrimnesBioimpedance}
	S. Grimnes, Bioimpedance. In: Wiley Encyclopedia of Biomedical Engineering 1. Wiley, 2006. 
	
	\bibitem{BMS2}
	S. Kaufmann, A. Malhotra, G. Ardelt and M. Ryschka, A high accuracy broadband measurement system for time resolved complex bioimpedance measurements, submitted to the IOP Journal of Physiological Measurements - Special Issue Electrical Impedance Tomography and Bioimpedance, 2014
	

	\bibitem{KuscheProjekt}	
	R. Kusche. Phasendetektor f\"ur die Bio-Impedanz-Messung, Fachhochschule L\"ubeck, Projektarbeit, 2012
	
	\bibitem{Halbleiterschaltungstechnik}
	U. Tietze, and C. Schenk, Halbleiterschaltungstechnik, 11th ed. Springer, 1999. 
	
	
	\bibitem{AhighaccuracyBMS}
	S. Kaufmann, G. Ardelt, and M. Ryschka, A high accuracy Bioimpedance Measurement System - System Design and first Measurements. Proceedings of the 5th International Workshop on Impedance Spectroscopy, 2012.
	
	\bibitem{MESI}
	S. Kaufmann, G. Ardelt, and M. Ryschka, Measurements of Electrode Skin Impedances using Carbon Rubber Electrodes - First Results, Journal of Physics: Conference Series 434(1), 2013.
	

	
\end{thebibliography}
\end{document}